\documentclass{report}
\usepackage[utf8]{inputenc}
\usepackage{amsmath, amssymb, graphics, geometry, graphicx}
\usepackage{amsthm}
\usepackage[
    backend=biber,
    style=apa,
  ]{biblatex}
\usepackage[colorlinks=false]{hyperref}
\usepackage{float}

\newtheorem{theorem}{Theorem}
\newtheorem{definition}{Definition}
\newtheorem{lemma}[theorem]{Lemma}

\newtheorem{example}{Example}[section]

\begin{document}

\title{Measuring Statistical Evidence: A Short Report}
\author{Mahdi Zamani \\ \\ Supervised by Professor Michael Evans \\ \\ \\ \\ University of Toronto \\ \\ }
\maketitle

\tableofcontents
\chapter{Statistical Problem}

\section{Introduction}
My interest in the field of Statistics was sparked by Hume's Problem of Induction \parencite{TreatiseHumanNature}. Since Hume introduced this 
problem in 1739, probability theory seems to be the most suitable framework for addressing it. While the mathematics of 
probability provides a \textit{somewhat} (due to Gödel) consistent system for reasoning, it does not prescribe how to interpret the resulting 
probabilities. Nevertheless, differences in interpretations of probability should not lead to divergent methods of 
statistical inference. I believe that the interpretation of probability is crucial for uncertainty quantification, but the 
rest of statistical inference should be conducted logically, and probability theory remains our best tool for tackling 
statistical problems.
The term "statistics" is derived from several European languages, including the Latin "status," the Italian "statistia," 
the German "statistik," and the French "statistique," all of which relate to a political state. Historically, statistics 
referred to information valuable to the state, such as data on population sizes (human, animal, products, etc.) and military 
strength.
An instance of an archetypal statistical problem is where there exists a finite population $\Omega$, and some measurements 
have been taken by $X:\Omega \to \mathcal{X}$. Then for a set $\mathcal{A} \subset \mathcal{X}$, the fundamental object of interest 
is the relative frequency ratio $\frac{\#(\{\omega \in \Omega : X(\omega) \in \mathcal{A}\})}{\#(\Omega)}$. 
So let the relative frequency function be $f_X(x) = \frac{\#(\{\omega \in \Omega : X(\omega) = x\})}{\#(\Omega)}$.
If conducting a census would be feasible, then there is no problem to be solved. However, in real life, doing so is typically
not possible or plausible. Hence the fundamental problem that statistics is trying to address is how to infer the true 
relative frequency function, from observing the measurements for a subset of the population. 
Like any other science, a first step is to impose some assumptions in order to be able to work with the problem. 
Although one can be skeptical about it, statisticians often assume that the data is coming from a certain family of relative 
frequency functions, where are indexed by the model parameter $\theta \in \Theta$. Hence, for parameter space $\Theta$, 
$\mathcal{M}  = \{f_\theta : \theta \in \Theta\}$ is our statistical model. The model can be largely in error, but imposing 
assumptions is part of science. The more important thing is the logical procedure that we are following which should be falsifiable 
and free of paradoxes. Moreover, there are issues that arise in statistics due to under-specification of the problem,
which will be ignored for the interest of this text. (See for instance the Borel paradox).
There are two base problems in statistics, namely estimation and hypothesis assessment where, naturally, an estimation should 
be supplied with a measure of accuracy (typically the "size" of a set which contains the estimate) and similarly for the hypothesis assessment. 
\subsection{Falsifiablity, Objectivity and Subjectivity}
Much effort in statistics seems to be in pursuit of reasoning objectively, and certain schools of thoughts such as 
frequentism had been developed with that hope. However, I believe that there is no way to avoid subjectivity in the process 
of statistical inference, or better say, science in general. When a statistician chooses a model, he is making a subjective 
choice, measurements that we take, Newton's $F=ma$ is also subjective etc. On the other hand, reaching the objective truth 
has always been the goal of scientific investigation. Statistics should be viewed as a way of reasoning where, although 
subjective choices are made along the way, choices can be checked with an (assumed) objective information. Hence, in 
our view, we assume that the data is generated objectively. Moreover, as mentioned in \parencite{popperLogicScientificDiscovery2002}, a valid scientific 
theory must be empirically testable; This is known as \textit{the falsifiability principle}. 
Hence, we believe that besides being logical, ingredients of a statistical inference method must be falsifiable. This 
view may not be accepted by Bayesians as they do not want to check priors. Parallel to Bayesians, certain aspects of frequentist inference 
such as using squared loss (in Mean Squared Error) are not falsifiable either.  
Certainly, one can question whether the data is chosen objectively or not, or is that even possible. As will be discussed shortly, there
is no way to be certain whether something is random or not. However, remembering that assumptions are a necessary part of 
science, we assume such a thing in our statistical analysis and require the collected data to be as objective as possible. 

\subsection{Randomness}
There have been various approaches in history to define and characterize randomness. So far, the most successful one seems to be
Kolmogorov's definition of randomness. There are several formal treatments of Kolmogorov complexity, which the interested
reader can refer to \parencite{liIntroductionKolmogorovComplexity2008}.
In essence, Kolmogorov complexity is the length of the shortest possible program (in any programming language), that can
generate a sequence of numbers. If the length of the sequence of numbers where greater or equal to its Kolmogorov 
complexity, then the sequence is called random. Therefore, in this treatment, probability does not play a role in randomness.
Moreover, it is worth noting that Kolmogorov complexity is not computable and more generally, there does not seem to be a 
way to identify if something is random or not. All of our statistical checks can only check if the given sequence possess
some desirable properties that a random sequence should have, but the converse does not hold. For instance, consider the 
Chambernowne's Sequence.
\begin{example}
 Suppose $\Omega = \{0, 1, 2, 3, 4, 5, 6, 7, 8, 9\}$ and assume the sequence: $0, 1, 2, 3, 4, 5, 6, 7, 8, 9, \\ 1, 0, 
 1, 1, 1, 2, 1, 3,\dots$ is generated. Clearly, this sequence is constructed deterministically from counting natural numbers
 in their order. However, it can be shown that the limiting relative frequency of any $\omega \in \Omega$ is $\frac{1}{10}$ 
 and the sequence is generated i.i.d from a uniform distribution. 
\end{example}

\subsection{Infinity and Continuity}
The basic statistical problem that we introduced had a finite sample space, and in the real world, I believe everything will be 
reduced to the finite case. However, for mathematical convenience, infinite sets can be used as an approximation to something
that is essentially finite. Indeed this simplification might introduce error in our analysis and certain care is needed. 
Moreover, there are different views regarding continuity in statistics. Some argue that continuity is a fundamental truth, 
but I believe that not only continuity arises as approximation, but also by taking it as a fundamental object, there are
various paradoxes that can arise. For instance, consider the following example inspired by Fisher's famous example where he used to incorrectly object
the Bayesian statistics \parencite{fisherMathematicalFoundationsTheoretical1922}. 
\begin{example}{Uniform Priors} \\
   Let $\Omega$ be our sample space of students with size $N$, where $N$ is very large. Suppose further that a measurement 
   will be taken $X:\Omega \to \{0, 1\}$ where for any $\omega \in \Omega$, 
   $$X(\omega) = \begin{cases}
   1, & \omega \text{ is a Statistics student} \\
   0, & \text{otherwise}
   \end{cases}$$ \\
    Hence our model for the observed data $x$, would be $\mathcal{M} = \{ \theta^{x}(1-\theta)^{1-x} : \theta \in \Theta_N\}$, where
    \\ $\Theta_N = \{0, \frac{1}{N},\dots\frac{N - 1}{N}, 1\}$.
    In Bayesian setting, which will be discussed later, a prior probability distribution will quantify our uncertainty about the 
    true value of the $\theta$. Let $\Pi_N$ be the density function of uniform probability distribution on $\Theta_N$. 
    Now since $N$ is very large, $\Theta_N$ can be approximated by $\Theta = [0, 1]$. Hence $\Pi(\frac{i}{N+1}, \frac{i+1}
    {N+1}) = \frac{1}{N+1}$. 
    Now assume that we want to do inference on a  1- 1 transformed parameter space via $\Psi: \Theta \to \varPsi$, where 
    $\Psi(\theta) = \theta^{2}$. This induces the prior probability density on $\varPsi$, $$p_{\Psi}(\psi) = \begin{cases}
    \frac{1}{2\sqrt{\psi}}, & \psi \in [0, 1] \\
    0, \text{otherwise}
    \end{cases}$$
    Some argue that this is a contradiction, since although $\Theta$ and $\varPsi$ are isomorphic, the probability distribution
    on $\varPsi$ is not uniform. This objection rises with the view of taking continuous models as fundamental object. 
    However, by considering continuity as an approximation, we know that the true value $\psi$ is in $\Psi_N = 
    \{0, \frac{1}{N^{2}},\dots \frac{(N-1)^2}{N^{2}}, 1\}$, and 
    $$
    \frac{1}{N+1} = \Pi(\frac{i}{N+1}, \frac{i+1}{N+1}) = \Pi_{\Psi}((\frac{i}{N+1})^2, (\frac{i+1}{N+1})^2) = 
    \int_{(\frac{i}{N+1})^2}^{(\frac{i+1}{N+1})^2} \frac{1}{2\sqrt{\psi}}d\psi
    $$
    Where $\frac{1}{2\sqrt{\psi}}$ is adjusting for the fact that the transformation is modifying length at different rates. 
    Hence there is no contradiction if we take the approximation into account. 
    \qed
\end{example}
For mathematical details of how probability density functions arise via a limit, see \parencite{evansMeasuringStatisticalEvidence2015}, Appendix. 
Moreover, an important point that the above example illustrates is the need for a meaningful discretization. Since we believe
that at the end of the day our inferences are for finite parameter spaces, a discretization $\delta$ must be supplied by 
the user for a given application. 

\subsection{Decision Theory}
Statistics can be partitioned into two school of thoughts throughout the years. Namely decision-theoretic (American) 
and evidential (British). 
Following the falsifiability principle, it is not clear how to check for ingredients in decision-theoric statistics such as 
loss function, utilities etc. Hence, this view is omitted in this text and although very interesting, it does not seem to 
help with constructing a logical and falsifiable inference methodology that can be used in scientific applications. Moreover,
it is worth mentioning that such treatments of statistics are not free of paradoxes, see for instance \parencite{Peterson_2017}.

\section{Probability}
Since probability lies at the heart of statistics, it needs to be discussed. We skip the history and mathematical 
details and rather focus on its various interpretations. In order to do so, we need to set some foundations. 

\section{Kolmogorov Formalization of Probability}
Throughout this text we assume $(\Omega, \mathcal{A}, \mathbb{P})$ is our probability triple, where $\Omega$ is our sample 
space, $\mathcal{A}$ is a sigma-algebra on $\Omega$ and $\mathbb{P}:\mathcal{A} \to [0, 1]$ is our probability measure such 
that $\mathbb{P}(\Omega) = 1$ and $\mathbb{P}$ is countably additive. 
Although, there has been attempts on working with finitely additive probability measure, since countable additivity implies 
continuity, its existence is necessary for conditional probability to behave correctly.

\section{Conditional Probability}
\begin{definition}{Principle of Conditional Probability} \\
   For $(\Omega, \mathcal{A}, \mathbb{P})$ and $A,C \in \mathcal{A}$ with $\mathbb{P}(C) > 0$, if it is known that event $C$ has 
   occurred, then $\mathbb{P}(A)$ must be replaced by $\mathbb{P}(A|C)$ 
\end{definition} 
Concerns has been raised for above principle in the philosophy literature as \textit{The Problem of Old Evidence} \parencite{glymourTheoryEvidence1980}. However, if the 
problem is characterized in statistical setting, the mentioned issue is resolved. 
It should be noted that we take this principle as an axiom. In other words, there is no mathematical justification of why we should
do so, but this seems the most plausible way to modify beliefs. 

Misapplication of the principle of conditional probability has created significant confusion in statistics community, such
as The Monte Hall Problem, The Prisoners Dilemma, etc. The root of this misunderstanding is the absence of a consistent way 
of conditioning on data. In order to address this, an Information Generator function $\Upsilon$ must be specified; 
$\Upsilon : \Omega \to \Xi$ is a function on the specified sample space $\Omega$, such that for a given context when $\xi_0 \in \Xi$ is true, the 
obtained information can be specified by $B = \Upsilon^{-1}\{\xi_0\}.$ 
\parencite{evansMeasuringStatisticalEvidence2015} Example 2.2.2 gives a detailed treatment of addressing The Prisoners 
Dilemma and The Monte Hall Problem.

\section{Subjective Probability}
A popular interpretation of probability is that it measures one's degree of belief about the occurrence of an event.
This idea does not have anything to do with objectivity. There are several justifications for subjective interpretation 
of probability such as probability via betting, scoring rules, Savage's axiomatization, Cox's theorem etc. Although very interesting and 
intelligent, usually there is one or two assumptions in these justifications that are very controversial, such as the 6th
axiom in Savage's axiomatization or 5th in Cox's. However, we believe that probability measures one's belief regardless of 
how it is assigned. 

\section{Relative Frequency Probability}
Contrary to the subjective probability camp, there are people who believe that probabilities correspond to real-world 
entities. Hence, in their view, 
an event's probability is its relative frequency in infinitely many trials. 
However, the existence of a random system for this 
definition seems essential and the corresponding issues about randomness have been discussed before. 
To me, while certainly useful in some applications, this interpretation 
looks far from reality and at best can be considered as a thought experiment.

\chapter{Survey Of Characterizing Statistical Evidence}
In this chapter we will be discussing important approaches that have been made in the literature
to characterize statistical evidence, and assess their shortcomings. 

\section{Pure Likelihood Inference}
Likelihood inferences are solely based on the likelihood function. 
\begin{definition}{Likelihood function}\\
For the observed data $x$ and model $\mathcal{M} = \{f_\theta : \theta \in \Theta\}$,
$\mathcal{L}(.|x): \Theta \to [0, \infty)$
is the likelihood function.
Where $\mathcal{L}(\theta|x)=kf_\theta(x)$ for some positive k.
\end{definition}

\subsection{Full Parameter Estimation}
Although the motivation comes from the discrete case as we argued is the case in real world, $\mathcal{L}$ 
imposes a preference ordering on $\Theta$ more generally.
If $\mathcal{L}(\theta_1|x) \leq \mathcal{L}(\theta_2|x)$, then $\theta_1$ is not preferred to 
$\theta_2$. i.e. $\theta_1 \preccurlyeq \theta_2$.
Note that in this case the $k$ gets cancelled and so if observing $x$ under $f_{\theta_1}$ is less probable than 
$f_{\theta_2}$, then $\theta_1$ is not preferred to $\theta_2$.
Naturally, parameter estimation arises from maximizing the likelihood function.
\begin{definition}{Maximum Likelihood Estimate (MLE)} 

$$ \theta_{MLE}(x) = \operatorname*{arg\,sup}_{\theta}\mathcal{L}(\theta|x)$$

\end{definition}
We assume that the MLE always uniquely exists but this is not always the case in the absence of sufficient amount of data. 
Moreover, there is another principle in pure likelihood approach regarding measuring the strength of evidence which follows
from the likelihood preference ordering on $\Theta$.
\begin{definition}{The Law of the Likelihood} \\
    For $\theta_1, \theta_2 \in \Theta, \mathcal{L}(\theta_1|x)/\mathcal{L}(\theta_2|x)$ measures the strength of the evidence supporting $\theta_1$
    over $\theta_2$.
\end{definition}
The above ratio is referred to as \textit{the relative likelihood}.

As argued in the previous chapter, it is natural to assess the accuracy of our estimate, MLE, with the \textit{"size"} of 
a set $\mathcal{C}(x) \subset \Theta$ where $\theta_{MLE}(x) \in \mathcal{C}(x)$. 
\begin{definition}{Likelihood Reigon}
   $$\mathcal{C}(x) = \{\theta:\mathcal{L}(\theta|x) \geq c(x)\}$$ 
   For some $c:\mathcal{X} \to [0, \infty)$.
\end{definition}
In accordance to the Law of the likelihood, and to avoid arbitrary choices for $c(x)$, we attempt to measure evidence 
on a \textit{universal} scale as follows: 
\begin{definition}{$(1 - \gamma)$-likelihood region for $\theta$ }
    
$$ C_\gamma(x) = \Biggl\{\theta:\frac{\mathcal{L}(\theta|x)}{\mathcal{L}(\theta_{MLE}(x)|x)} \geq 
\frac{c(x)}{\mathcal{L}(\theta_{MLE}(x)|x)} = 1 - \gamma\Biggl\}$$
for some specified $\gamma \in [0, 1]$.
\end{definition}
Hence, for a given $\gamma$, $C(x)$ contains values $\theta \in \Theta$ such that the data supports $\theta$ at least $100(1 - \gamma)\%$
of the maximum support (MLE). 
However, this approach still needs to provide a guide for choosing $\gamma$. \parencite{royallStatisticalEvidenceLikelihood2017} does so by arguing based on an
urn model that for $\theta$, whenever the relative likelihood $\mathcal{L}(\theta|x) / 
\mathcal{L}(\theta_{MLE}|x) \geq \frac{1}{8}$ then there's strong evidence in support of $\theta$.
As a result, same logic can be applied for hypothesis testing as well. 
Personally, I have not looked into Royall's urn model argument closely but I believe there is no sound justification 
for $\gamma$. While besides Royall's urn model argument likelihood inference seems to be uncontroversial, there are problems 
associated with whether or not the relative likelihood ratio is measuring strength of evidence. To illustrate this point,
consider the following discrete example in \parencite{evansMeasuringStatisticalEvidence2015}, to control for the issues that might arise due to infinity.

\begin{example}

Let $\mathcal{A} = \{a_1, a_2,...,a_k\}$ be set of letters, and $\Theta_k$ be the set of all words of length $M$ or less.
Define $l : \Theta_k \to \mathbb{N}$ to measure the size of a word in $\Theta_k$, and $r : \Theta_k \to \Theta_k$ s.t 
$r(\theta)$ is $\theta$ with the last letter chopped.
Now, for this inference problem, let $\Theta_k = \mathcal{X}_k$ and suppose $x$ is observed. Let $\delta > 0$ and define
the probability distribution as follows : \\
If $l(\theta) < M$ 
$$
f_{\theta}(x) = \begin{cases}
    \frac{1}{k+1} + \delta, & x=\theta \\
    \frac{1}{k+1} - \frac{\delta}{k}, & x=\theta a_i \text{ for } i=1,...k \\
    0, & \text{o/w}
\end{cases}
$$
If $l(\theta) = M$ 
$$
f_{\theta}(x) = \begin{cases}
    1, & x=\theta \\
    0, & \text{Otherwise}
\end{cases}
$$
By using above, the relative likelihood is : 
$$
\frac{\mathcal{L}(\theta|x)}{\mathcal{L}(\theta_{MLE}(x)|x)} = \begin{cases}
    1, & \theta=x \\
    \frac{1}{k+1} - \frac{\delta}{k}, & \theta=r(x) \\
    0, & \text{o/w}
\end{cases}
$$

Notice that for a small value of $\delta$ and by choosing $k$ large enough, $\mathcal{L}(\theta|x) / 
\mathcal{L}(\theta_{MLE}(x)|x)$ can be made arbitrary small when $\theta=r(x)$. Hence for any $\gamma < 1$,
we can construct $\mathcal{C}_\gamma(x) = \{x\}$. \\ 
So we have a very high accuracy for MLE 
and on the other hand, for $l(\theta) > 0$, $\mathbb{P}_\theta(\theta=r(x)) = \frac{k}{k+1} - \delta$. Note that by proper 
choice of $k$, we can make $\mathbb{P}_\theta(\theta=r(x))$ arbitrary close to $1 - \delta$. 
As a result, for many observed values $x$, our MLE estimation has a very high accuracy but we are virtually certain that
the true value is $r(x)$. For more information on another variation of this example, take a look at \parencite{evansExampleConcerningLikelihood1989}.
\\
\qed
\end{example}
As the example illustrated, there are concerns with likelihood as this does not seem to measure the strength of 
evidence in favor of $\theta$ and this raises concerns regarding the calibration in the relative likelihood. 

\subsection{Marginal Parameter Estimation}
Consider the usual setup of a statistical problem. Furthermore let $\Psi : \Theta \to \Xi$ \textit{not} be a 1-1 function. 
Assume that we are interested in assessing a composite hypothesis. i.e. $H_0 = \Psi^{-1}\{\xi\} \subset \Theta$. 
Since the pure likelihood approach is silent in this setting, \textit{profile likelihood} has been introduced to resolve
the issue. 
\begin{definition}{Profile Likelihood Function}
    $$
        \mathcal{L}^{\Psi}(\xi|x) = \operatorname*{sup}_{\theta \in \Psi^{-1}\{\xi\}} \mathcal{L}(\theta|x)
    $$
\end{definition}
Similar to likelihood function, this induces a preference ordering on $\Xi$. Moreover, it can be shown that, under weak conditions, $\gamma$-profile 
likelihood region is invariant under reparameterization. Namely, projecting $\mathcal{C}_{\gamma}(x)$ via a one-to-one mapping $\Psi$ yields
the same inference.
However, the issue with this approach is that $\mathcal{L}^{\Psi}(.|x)$ is not a likelihood function in general. So the 
profile likelihood method needs a justification itself which is out of the scope of the pure likelihood principle. See \parencite{evansMeasuringStatisticalEvidence2015} Example 3.2.2
for an instance where the profile likelihood function is not a likelihood function.
Moreover, other forms of likelihood functions such as \textit{integrated likelihood}, \textit{marginal likelihood} and \textit{
conditional likelihood} has been developed, but they only work in limited contexts and cannot be applied in general.
To conclude, the pure likelihood inference and its variants suffer various issues and the root of that seems to be the effort to measure the 
evidence on a universal scale. 

\section{Birnbaum's Theorem}
Birnbaum has made considerable contributions to the literature on statistical evidence. A controversial result obtained in 
\parencite{birnbaumFoundationsStatisticalInference1962} was that by accepting two frequentist principles, namely sufficiency and conditionality, one 
must adhere to the likelihood principle as they are equivalent. Unfortunately this is largely ignored in today's 
statistics courses but it is indeed important, because this indicates that important tools in frequentist inference such as 
p-values, confidence regions, repeated sampling, etc will be left out. In order to see what Birnbaum did (and did not), 
we need a useful formalization for characterizing statistical inference. 

Define an inference base $I = (\mathcal{X}, \mathcal{M}=\{f_\theta:\theta \in \Theta\}, x)$, where $\mathcal{X}$ is the 
sample space, $\mathcal{M}$ is the model and $x \in \mathcal{X}$ is the observed data. Let $\mathbf{I}$ be the set of 
all such inference bases. Then, 
\begin{definition}{Statistical Principle} \\
   Whenever $R \subset \mathbf{I} \times \mathbf{I}$ is an equivalence relation, it is called a \textit{Statistical 
   Principle}. 
\end{definition}
The underlying idea is that if two inference bases are related by some principle $P$, then they contain the same 
statistical evidence regarding inferring the true parameter under that principle.  \\
For the sake of notation, if $R$ is a relation on set $D$,
then the equivalence relation $\bar{R}$ generated by $R$ is the smallest equivalence relation containing $R$. It is worth
noting that $\bar{R}$ might not always happen to be meaningful in statistical setting and it should be examined.

Consider the following statistical principles. 
\begin{definition}{Likelihood Principle ($L$)} \\
Let  $I_0 = (\mathcal{X}, \{f_{0\theta}:\theta \in \Theta\}, x_0)$ and $I_1 = (\mathcal{X}, \{f_{1\theta}:\theta \in 
\Theta\},x_1 )$
be inference bases and define $L \subset \mathbf{I} \times \mathbf{I}$ by 
 $(I_0, I_1) \in L$ whenever for every $\theta \in \Theta$, there exists $c> 0$ s.t $f_{0\theta}(x_0) = cf_{1\theta}(x_1)$ 
\end{definition}

The following lemma is needed for defining the sufficiency principle. 
\begin{lemma}{(Minimal) Sufficient Statistic $T$} \\
  A function $T$ is a sufficient statistic whenever the conditional distribution of $x$ given $T(x)$ is independent
  of $\theta$. In other words, $T$ contains all the needed information for inference. 
  $T$ is said to be \textit{minimal}, whenever for any other sufficient statistic $U$, there exists a function $h$ such 
  that $T = h \circ U$ 
\end{lemma}

\begin{definition}{Sufficiency Principle ($S$)} \\ 
  Let $T_0$ be the minimal sufficient statistic for model $\mathcal{M}_0$ and $T_1$ be such for $\mathcal{M}_1$. Moreover,
 let $\mathcal{M}_{0,T_0}$ and $\mathcal{M}_{1,T_1}$ be the respective marginal models. 
 $(I_0, I_1) \in S$ whenever there exists a 1-1 function $h$ between the sample spaces of the marginal models, and 
 $T_0(x_0) = h(T_1(x_1))$.  
\end{definition}
Before defining the last \textit{relation}, we need to following definition. 
\begin{definition}{Ancillary Statistic} \\
    A function $h$ on $\mathcal{X}$ is an ancillary statistic for model $\mathcal{M}$ if the distribution
    of $h$ is independent of $\theta \in \Theta$. Hence the value of $h(x)$ is silent about the true value of $\theta$. 
  Moreover, for $x \in \mathcal{X}$, the conditional model given $h(x)$ is $\mathcal{M}'=\{f_{\theta}(.|A(x)):\theta \in 
  \Theta\}$.
\end{definition}

For a motivation of the following relation, see \parencite{evansMeasuringStatisticalEvidence2015}, Example 3.3.1.
\begin{definition}{Conditionality \textit{Relation} ($C$)} \\
 $(I_0, I_1) \in C$ whenever sample spaces and observed data of both inference bases are equivalent and there exists an 
 ancillary statistic for $\mathcal{M}_0$, $h$, such that the conditional model given $h(x_0)$ is $\mathcal{M}_1$ or 
 the same holds for $(I_1, I_0)$. \\
\end{definition}
Additionally, it is worth noting that $S \subset L$ and $C \subset L$
and it must be emphasized that $C$ is not an equivalence relation and hence can't be a statistical principle. This is because 
$C$ is not transitive.
As shown in \parencite{evans2013doesproofbirnbaumstheorem}, Birnbaum's theorem was not correctly stated. In fact, Birnbaum's theorem establishes the following 
result. 
\begin{theorem}{Birnbaum's Theorem}
   $$
   S \cup C \subset L \subset \overline{S \cup C}
   $$ 
\end{theorem}
As shown in \parencite{evansMeasuringStatisticalEvidence2015}, $C \subset L$ and $\overline{C} = L$. Hence accepting the relation $C$ does not necessarily lead to accepting
$L$ unless the extra elements, $\overline{C} \setminus C$, make sense. Moreover, it is established in \parencite{evansMeasuringStatisticalEvidence2015} that 
$L = \overline{S \cup C}$. Thus, Birnbaum is not proving what is has been claimed over the years, but rather it is 
showing that the $L$ is the smallest equivalence relation that contains $S \cup C$. i.e $\overline{S \cup C}$. As a result, 
Birnbaum's theorem does not provide support for the likelihood principle.

\section{Frequentist Approach: P-values}
P-values play a central role in the frequntist approach regarding evidence. As defined in \parencite{coxTheoreticalStatistics1979}, Let 
$\mathcal{M} = \{f_\theta : \theta \in \Theta \}$ be our model, $x$ the observed data, $H_0 \subset \Theta$ be the 
null hypothesis, and $T:\mathcal{X} \to  \mathbb{R}$ be our test statistic. Let $\mathbb{P}_{H_0, T}$ be the
marginal distribution of $T$ under $H_0$, which is fixed for any $\theta \in H_0$. Then, 
\begin{definition}{P-value} \\ 
   $$
    p_{H_0}(x) = \mathbb{P}_{H_0}(T(X) \geq T(x))
   $$
\end{definition}
This definition corresponds to the term "p-value", as the "p" stands for probability. 
In the frequntist perspective, a small value of $p_{H_0}(x)$ is considered as evidence against the null hypothesis. 

However, there are several issues regarding p-values that the interested
reader can trace via the term \textit{p-hacking} in the literature. Here we restrict our attention to some of the important issues
with p-values. \\  
First thing to note is that the same issue from pure likelihood theory carries over; Namely, we need to specify a cut-off
, $\alpha$, where we can decide whether there is evidence against our null hypothesis or not. Empirically and traditionally
$\alpha$ is often set to $0.05$, but this is really context dependant and the issue for choosing $\alpha$ is not resolved
so far. 

Moreover, an important problem with the p-value is that it is not sensitive to sample size. To illustrate this point, consider
the following example. 
\begin{example}{Location Normal} \\
   Let $\mathcal{M} = \{f_\theta : \theta \in \Theta \}$ be our model, where $f_\theta \sim N(\theta, 1)$ 
   and let $T(X) = \bar{X}_n$. Then $T(X) \sim N(\theta, \frac{1}{n^{2}})$, and as argued in chapter 1, we specify
   a $\delta$ as the precision of our measurement. 
   It can be easily seen that the more data we collect, the more the probability distribution of $T(X)$ will be 
   concentrated around the null hypothesis, when then null hypothesis is true (Figure~\ref{locnormal}). This concentration can be done arbitrarily more as much as putting the 
   probability mass virtually within $\delta$ distance from null. In this case, we have overwhelming evidence in favor 
   of the null, but the p-value, since it is not sensitive to sample size, might still suggest to reject the null. \qed
   \counterwithout{figure}{section}
   \begin{figure}[H]
    \centering
    \includegraphics[scale=0.45]{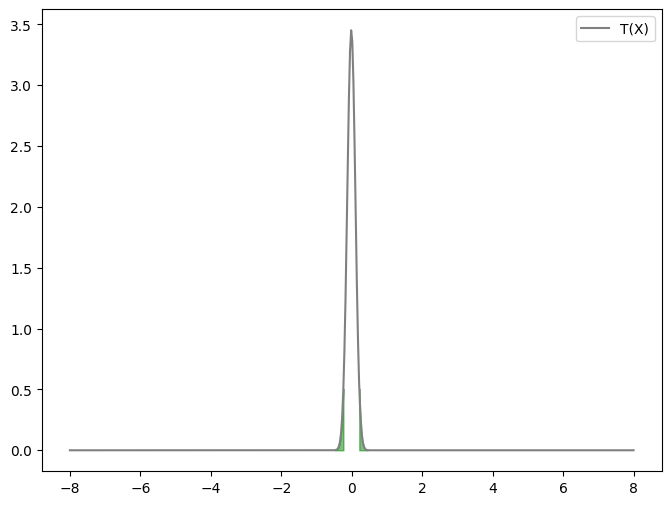}
    \caption{$H_0 : \theta = 0$}
    \label{locnormal}
\end{figure}
\end{example} 
It is widely and wrongly believed that the large values of $p_{H_0}(x)$ indicates evidence in favor of the null. However, since the p-value is 
distributed uniformly under the null (at least in continuous case), no particular value of $p_{H_0}(x)$ is expected and hence it cannot prescribe
whether the evidence is in favor of the null. This is a huge down side of p-values, as it is expected from any valid measure of evidence to able to 
prescribe whether there is evidence in favor of the null.

Moreover, in the setting where the data is collected sequentially, one can argue that the assessment of
evidence should be independent of the stopping rule. However, p-values suffer from such issue. Imagine a scientist with a 
determined $\alpha$, collects $n$  samples and obtains a p-value of $\alpha + \epsilon$, where $\epsilon > 0$. Thus, with 
a hope of rejecting the null, he collects $m$ more data. 
Let $A$ be the event that the evidence has been found against the null at the first stage, and $B$ be the event that such
happened in the second stage. Then, considering $n + m$ data, the probability of finding evidence against the null in the
first stage or the probability of not finding evidence against the null in the first stage and finding evidence against
the null in the second stage, will be more than $\alpha$. i.e $\mathbb{P}_{H_0}(A) + \mathbb{P}_{H_0}(A^{c} \cap 
B) > \alpha.$ 
Therefore, it is not possible to find evidence against the null! There are other measures of evidence, such as \textit{e-values}, that does not suffer this issue.
However, they share lots of similarities and hence some concerning issues are similar as well. This is briefly discussed in \parencite{evans2024conceptstatisticalevidencehistorical} section 3.1 and the reader should 
consult \parencite{ramdas2024hypothesistestingevalues} for a comprehensive treatment of e-values.

The above examples illustrate some serious issues with p-values, and I believe this is sufficient to conclude that p-values 
are not a valid measure of evidence. Moreover, it is worth emphasizing that there is no definitive guide for finding the right 
test statistic and this ambiguity can result in totally different inferences. An example of this is illustrated in 
\parencite{evansMeasuringStatisticalEvidence2015} Example 3.4.2. 

\section{Frequentist Approach: Confidence Intervals}

P-values are deeply related to confidence intervals. Suppose for every $\theta \in \Theta$, there is a test statistic 
$T_\theta$ and as argued above, there exists a p-value function $p_\theta(x)$. Then, a $(1 - \alpha)$-confidence region
is defined as follows:
\begin{definition}{$(1-\alpha)$-Confidence Region} is defined as
$\mathcal{C}_{1-\alpha}(x) = \{\theta : p_\theta(x) > \alpha\}$
\end{definition}
Note that $\mathbb{P}_\theta(\theta \in \mathcal{C}_{1-\alpha}) \geq 1 - \alpha$.
By the above definition, all the problems associated with p-values carry over to confidence intervals as well. 
A somewhat strange procedure in frequntist inference, to the contrary of pure likelihood theory, is that after 
determining the $\mathcal{C}_{1-\alpha}(x)$, we still need to figure out our estimation of the parameter of interest. 
This is odd, because the purpose of confidence regions are to assess the accuracy of the estimation. 
Moreover, it is not always possible to construct meaningful confidence regions. For instance,
\begin{example}
    Assume $\Theta = [0, 1]$ and $x \in \mathbb{R}$. 
    Let $\mathcal{M} = \{f_\theta : \theta \in \Theta\}$ be our model, where $f_\theta(x) = ( 1 - \theta)\varphi(x)
    + \theta \varphi(x - 1)$ 
    and $\varphi$ is the density function of the $N(0, 1)$ 
    Then, 
    $$
    \mathcal{C}(x) = \begin{cases}
        [0, 1], & -1.68148 \leq x \leq 2.68148 \\
        \phi , & Otherwise
    \end{cases}
    $$
    Clearly, since $\Theta = [0, 1]$, this is not informative. Moreover, it is worth noting that this is an unbiased and
    uniformly most accurate confidence region! \\
    \qed 
\end{example}

\section{Bayesian Inference}
Remembering the initial discussions about subjectivity and objectivity, Bayesian inference builds on the idea that the
statistician should provide a \textit{prior} probability measure on the parameter space as well as specifying the model. 
Then, the rest of the inference carries by using probability theory and the axiom of conditional probability as the
proper way to change one's beliefs. 
Let $\mathcal{M} = \{f_\theta : \theta \in \Theta\}$ be our model, $\Pi$ be the prior probability measure on 
$\Theta$ and $\pi$ denote the prior probability density function w.r.t to volume measure $v$ on $\Theta$. 
By choosing a model $\mathcal{M}$ and a prior probability measure $\Pi$ we can characterize the joint uncertainty for 
$(\theta, x)$.
Within this community, there are several approaches such as Quantile-based inference, Loss-based inference, Empirical 
Bayes, Hierarchical Bayes, Bayesian Frequentism, etc. We will be focusing on two important topics, namely MAP-based 
Inference and Bayes factors, as these will bring up important issues that will help us move towards a more ideal 
inference method. 

\subsection{MAP-based Inference}
MAP-based inference, as the name suggests, aims to maximize a posterior. As a result, a preference ordering is induced 
on $\Theta$ as follows: for $\theta_1, \theta_2 \in \Theta$, if $\pi(\theta_1|x) \leq \pi(\theta_2|x)$ then 
$\theta_1 \preccurlyeq \theta_2$. Hence, it naturally follows to maximize the posterior for estimation. 
\begin{definition}{MAP Estimate} \\ 
   $$\theta_{MAP}(x) = \operatorname*{arg\,sup}_{\theta \in \Theta}\pi(\theta|x)$$

\end{definition}
There are at least two issues associated with this type of inference. The subtle one is that MAP-based inference is not 
invariant under 1-1 reparameterizations for continuous parameter spaces. 
Let $\psi : \Theta \to \Xi$ be 1-1 and smooth. Then, the density of $\xi \in \Xi$ is: 
$\pi_{\psi}(\xi|x) = \pi(\psi^{-1}(\xi)|x) J_{\psi}(\psi^{-1}(\xi))$ w.r.t the volume measure on $\Xi$. 
Hence, whenever $J_{\psi}(\psi^{-1}(\xi))$ is not a constant function of $\xi$, it is possible that $\xi_{MAP}(x) \neq 
\psi (\theta_{MAP}(x))$. However, this is a subtle issue because of the discussion on infinity in the first chapter. In a
given application, this issue will resolve by discretizing the parameter space meaningfully.  

Perhaps the more fundamental issue with MAP-based inferences is the fact that by the induced preference ordering, we are 
measuring evidence solely based on the value of the posterior density function. Remembering the Bayesian inference 
setting, a large value of $\pi(\theta|x)$ can be caused by a large $\pi(\theta)$ and thus $\theta$ might not be
the true value. This can be the case in the absence of sufficient amount of data. 

\subsection{Bayes Factor}
Perhaps the most commonly used measure of evidence is the \textit{Bayes factor}. Here is the definition:
\begin{definition}{Bayes Factor} \\
   Let $A \subset \Theta$, $0 < \Pi(A) < 1$, and $x$ be observed. Then the Bayes factor in favor of $A$ is 
   $$
    BF(A|x) = \frac{\Pi(A|x) \Pi(A^{c})}{\Pi(A^{c}|x) \Pi(A)} = \frac{Odds(A|x)}{Odds(A)}
   $$
\end{definition}
Hence, $BF(A|x)$ is measuring the change of belief in terms of odds in favor. In this setting, $BF(A|x) > 1$ indicates evidence 
in favor of A, $BF(A|x) < 1$ indicates evidence against the statement that the true parameter is in $A$, and $BF(A|x) = 1$ 
is interpreted as there is no evidence against or in favor of $A$ containing the true parameter. Moreover, it is presumed
that the value of the Bayes factor determines the strength of the evidence. For instance, the larger the $BF(A|x)$, 
it is claimed that the more evidence there is in favor of $A$. In other words it indicates stronger evidence in favor of $A$. Jefferys came up with a scale where a Bayes factor greater than $100$ 
is decisive, between $10^{\frac{3}{2}}$ to $100$ is very strong, etc.

In order to dive deeper into concerns around Bayes factors, the following lemma is useful. 
\begin{lemma}
    Let $A \subset \Theta$ ,$0 < \Pi(A) < 1$, and $x$ be observed. Moreover, let $T$ be a minimal sufficent statistic for
    $\mathcal{M} = \{f_\theta : \theta \in \Theta\}$. Then
    $$
    BF(A|x) = \frac{m(x|A)}{m(x|A^{c})} = \frac{m_{T}(T(x)|A)}{m_{T}(T(x)|A^{c})}
    $$ 
    Where, $m(.|A)$ is the predictive prior density conditioned on $A$, and $m_{T}(.|A)$ is the prior predictive of $T$ 
    conditioned on $A$.
\end{lemma}
Consider the hypothesis testing setting, where we want to test whether the true $\theta$ is in $H_0 \subset \Theta$. 
An apparent concern about Bayes factors raises when $\Pi(H_0) = 0$, which can happen in real world applications, as then the Bayes factor is undefined.
\parencite{jeffreysTheoryProbability1998} proposes a solution by specifying a prior 
probability for $H_0$, and two conditional prior probability measure for $\Theta$, $\Pi(.|H_0)$ and $\Pi(.|H^{c})$. 
Then the prior is taken to be $\Pi'(A) = p\Pi(A|H_0) + 
(1 - p)\Pi(A|H_0^{c})$ for $A \subset \Theta$. I believe this is an instance of taking a solely mathematical approach 
towards statistics. Although by specifying such a prior, the Bayes factor will be defined, doing so is meaningless and 
seems very arbitrary. 
A more serious famous issue is discussed in the following classical example. 
\begin{example}{The Jefferys-Lindley Paradox} \\ 
 Let $\Theta = \mathbb{R}$ and $\mathcal{M} = \{f_\theta : \theta \in \Theta\}$, where $f_\theta$ is the densify function
 for $N(\theta, 1)$. Moreover, let $x = (x_1,\dots x_n) \in \mathcal{X}$ be observed i.i.d and $T(X) = \bar{X}$ 
 be a minimum sufficient statistic. Hence, $T(X) \sim N(\theta, \frac{1}{n^2})$. 

   Our goal is to assess the hypothesis $H_0 = \{0\}$.  
   For $A \subset \Theta$, $p > 0$ and $\sigma \in \mathbb{R}$, define the prior as $\Pi'(A) = p \delta_0(A) + (1-p)\varphi(A)$
   , where $\delta_0$ is the Dirac measure on $0$, and $\varphi(A) = \int_{A}^{}f(x)$ when $f(x)$ is the density function 
   for $N(0, \sigma^{2})$. 
   By working out the calculations, $m_{T}(T(x)|H_0)$ is the probability density function of $N(0, 1)$, and under $H_0^{c}$, $m_{T}(T(x)
   |H_0^{c})$ is the probability density of $N(0, 1+n\sigma^{2})$ evaluated at $T(x)$. Thus, by the Lemma, 
   $$
   BF(H_0|x) = \frac{m_{T}(T(x)|H_0)}{m_{T}(T(x)|H_0^{c})} = e^{- \frac{(n\sigma\bar{x})^2}{2(1+n\sigma^2)}}\sqrt{1 + n\sigma^{2}}
   $$ 
   Now by fixing $\bar{x}\sqrt{n}$, the Bayes factor approaches infinity as $\sigma \to \infty$. Hence, by opting for a 
   diffuse prior, we are inherently inducing more evidence in favor of $H_0$.\
   
   On the other hand, by using Frequentist
   approach, and setting $\bar{x}\sqrt{n} = 5$, the p-value will be $6 \times 10^{-7}$. This gives overwhelming evidence 
   against the null and is surprising, since in general it is perceived that by using diffuse priors, Bayesian and 
   frequentist inferences should lead to the same result. 
  Taking a closer look, The Bayes factor seems to behave well as a measure of evidence. This is because as $\sigma \to \infty$  
 , the bulk of the prior probability moves away from the null and then $\bar{x}\sqrt{n}$ looks more reasonable as a value 
  from a $N(0, 1)$. The root of the paradox is the fact that the value of Bayes factor does not measure the strength of 
  the evidence, but we leave the paradox open for now and will address it in the next chapter. 
   The other issue worth noting is that it is not clear how to choose $\sigma$ as the hyperparameter. Hence, this deviates
   from the logical procedure that a statistical inference methodology should possess.  
   \\ \qed
\end{example} 

\chapter{Relative Belief}
After assessing some of the currently used methods for measuring and drawing inference based on evidence, this chapter 
will introduce a methodology to measure statistical evidence, using \textit{Relative Belief Ratio}. In doing so, we will try 
to address some of the previously raised concerns in chapter 2. 

\section{Introduction}
Considering flaws of inference methods that do not explicitly define what evidence is, we will define explicitly what do we mean 
by evidence. Firstly, we note that evidence and belief are different; Probability measures degree of belief, and 
evidence is measured by the \textit{change} in belief. This distinction is very important, as probability merely captures
our belief about a certain event, which is subjective. However, assuming the data is objective, it is the data that 
results in the updated belief. Thus, measuring the change in belief seems to be the right way of defining what statistical
evidence is. 
From now on, consider $(\mathbb{P}, \mathcal{A}, \Omega)$ to be our probability triple model, and assume there is a valid
information generator. Furthermore, let $\omega \in C$ be the (validly) obtained information and $\mathbb{P}(C) > 0$. Then, 
\begin{definition}{Principle of Evidence} \\ 
   If $\mathbb{P}(A|C) > \mathbb{P}(A)$, then there is evidence in favor of event $A$ being true, and if 
    $\mathbb{P}(A|C) < \mathbb{P}(A)$, then there is evidence against of $A$ being true. Lastly, whenever 
    $\mathbb{P}(A|C) = \mathbb{P}(A)$, the data is not indicating evidence in favor or against $A$ being true.
\end{definition}
As argued, the proper way of measuring statistical evidence is through the change in belief and this can be done in 
multiple ways. Throughout history, there have been attempts to do so. \parencite{carnapLogicalFoundationsProbability1950} proposed possibly the simplest way Of
measuring evidence by $\mathcal{D}(C, A) = \mathbb{P}(A|C) - \mathbb{P}(A)$. However, this does not behave properly in the
continuous case. Perhaps the second most easy and natural method is to consider the ratio and this is indeed what we mean
by the \textit{relative belief ratio}. 
\begin{definition}{Relative Belief Ratio (simple case)} \\ 
    Provided that the concealed $\omega$ is in $C \subset \Omega$, relative belief ratio for $A \subset \Omega$ is
    $$
    RB(A|C) = \frac{\mathbb{P}(A|C)}{\mathbb{P}(C)}
    $$
\end{definition}
There is an axiomatic construction of relative belief which can be found in \parencite{evansMeasuringStatisticalEvidence2015}, but like any other axiomatization,
concerns can be raised. However, it's really the power that the theory gives us that demonstrates its appropriateness. It 
is worth mentioning that other famous proposed measures of evidence are 1-1 increasing functions of relative belief ration 
except for the Bayes factor which will be discussed individually.
There are a number of nice properties that are associated with this simple and intuitive definition, and we will touch
on a few. An important lemma is the following. 
\begin{lemma}{Savage-Dicky Ratio}\\
    Assume $\mathbb{P}(A), \mathbb{P}(C) > 0$. Then $$RB(A|C) = RB(C|A)$$
\end{lemma}
Perhaps the most interesting property is the general additivity. 
   If $\mathbb{P}(A \cap B) > 0$, then 
   $$
   RB(A \cup B|C) = RB(A|C)\mathbb{P}(A|A\cup B) +RB(B|C)\mathbb{P}(B|A\cup B) - RB(A\cap B|C)\mathbb{P}(A\cap B | A \cup
   B)
   $$
   This implies that whenever $A \cap B = \phi$ and $\mathbb{P}(B) > 0$, 
   $$
   RB(A \cup B|C) = RB(A|C)\mathbb{P}(A|A\cup B) +RB(B|C)\mathbb{P}(B|A\cup B)
   $$
For $A \subset B$, at first glance, it might not look plausible for $RB(A|C) > RB(B|C)$ to be possible. However, the above 
propery indicates that the evidence that $A$ is true is contributing to the evidence that $B$ is true, by the factor of 
the conditional probability $\mathbb{P}(A|B)$. Consider the following clarifying example. 

\begin{example}{Evidence of a crime} \\ 
  Suppose that a murder is committed in Toronto and it is known that the murderer is from Toronto. Let $m$ be the 
  population of Toronto. Suppose that it has been told with certainty that the murderer comes from neighbourhood $\alpha$; 
  Let $C$ denote this evidence.
  Also, there are $m_1 < m$ people of that neighbourhood in Toronto and assume there are $n$ university students in the town,
  which $n_1 < n$ of them are of neighbourhood $\alpha$. 
  Let $B$ be the event that a university of student commited the crime, and $A$ denote the event that a university 
  student of neighbourhood $\alpha$ has committed the crime. Now consider the following relative belief ratios. 
  $RB(A|C) = \frac{\mathbb{P}(A|C)}{\mathbb{P}(A)} = \frac{m}{m_1}$ and 
  $RB(B|C) = \frac{\mathbb{P}(B|C)}{\mathbb{P}(A)} = \frac{n_1 m}{m_1 n}.$
  Certainly, $RB(A|C) > 1$ and so there is evidence in favor of the event that a university student of neighbourhood $\alpha$ has committed the crime. 
  On the other hand, by proper choice of $\frac{n_1}{n}$, $RB(B|C)$ can be made less than $1$. 
  Hence, there is evidence in favor of the event that a university student of neighbourhood $\alpha$ has committed the crime,
  but there is evidence against the statement that a university student committed the crime. This is due to the ratio 
  of students of neighbourhood $\alpha$. i.e. $\frac{n_1}{n}$. This seems fair, since when a small fraction of university 
  students are from neighbourhood $\alpha$, it is not fair to claim that the evidence suggests that a university student 
  has committed the crime. 
\end{example}
There are other nice and simple properties about relative belief which can come almost for free, due to the 
definition of relative belief, but we skip here for the interest of this text.

In order to expand the usage of relative belief in the continuous case where the probability of an event might be zero, we define the generalized version of the relative belief ratio. 
\begin{definition}{Relative Belief Ratio (Generalized)} \\ 
  Let $(P, \mathcal{A}, \Omega)$ be our probability triple and let $f$ be the corresponding density function w.r.t 
  volume measure $v$ on $\Omega$. Suppose that $\psi : \Omega \to \Xi$ is smooth, where $\psi(\omega) = (\psi_1(\omega), 
  \psi_2(\omega))$. Moreover, let $f_\psi$ be its density function w.r.t the volume measure on $\Xi$. 
  For $\psi(\omega) = (\xi_1, \xi_2)$,
  $$
  RB_{\psi_1}(\xi_1|\xi_2) = \lim_{\delta, \epsilon \to 0} RB(N_{\psi_1, \delta}(\xi_1)| N_{\psi_2, \epsilon}(\xi_2))
  $$
  Also, under regularity conditions, 
  $$
  RB_{\psi_1}(\xi_1|\xi_2) = \frac{f_{\psi}(\xi_1 | \xi_2)}{f_{\psi_1}(\xi_1)}
  $$
  Where, $f_{\psi_1}$ is the marginal density of $\psi_1$ and $f_{\psi}$ is the conditional density of $\xi_1$, given 
  $\xi_2$. Moreover, 
  $N_{\psi_1, \delta}(\xi_1), N_{\psi_2, \epsilon}(\xi_2)$ are "nice" neighbourhoods. (for details on the 
  convergence and regularity conditions, see \parencite{evansMeasuringStatisticalEvidence2015} Appendix.)
\end{definition}
This corresponds to our view that continuity arises as a tool to approximate something that is essentially finite. 
In general, when $f_{\psi_1}(\xi_1) > 0$, the second formulation of relative belief can be employed, but it is essential 
to remember that ultimately the definition arises as a limit. 
The following utilizes relative belief ratio in Bayesian context. 
\begin{definition}{Bayesian Relative Belief Ratio} \\
   Let $\Omega = \Theta \times \mathcal{X}$ and $f(\theta, x) = \pi(\theta)f_{\theta}(x)$ be the density function. 
   Let $\Psi : \Theta \to \varPsi$, $\Upsilon: \Omega \to \Theta$ and $\Upsilon: \Omega \to \mathcal{X}$.
   Suppose further that $(\psi, x) = (\Upsilon_1(\theta, x), \Upsilon_2(\theta, x))$ where $\Upsilon_1$ doesn't depend on 
   $x$ and $\Upsilon_2$ is just a projection on $x$. Then, when $\pi_{\Psi}(\psi) > 0$, 
   $$
   RB_{\Psi}(\psi|x) = \frac{\pi_{\Psi}(\psi|x)}{\pi_{\Psi}(\psi)}
   $$
   Where, $\pi_{\Psi}(.|x)$ is the posterior density and $\pi_{\Psi}$ is the prior density.
\end{definition}
As argued in previous chapters, a desirable property of a measure of evidence it to be invariant under 1-1 
reparameterizations, and indeed relative belief possesses that property. 
\begin{theorem}{Invariance of Relative Belief} \\
    Assume $\Upsilon: \varPsi \to \Lambda$ is a smooth 1-1 transformation such that $\Upsilon(\psi) = \lambda$, then 
    $$
    RB_{\Psi}(\psi|x) = RB_{\Upsilon}(\lambda|x)
    $$
\end{theorem}

A number of earlier stated properties can get generalized. Here's the general version of the 
additivity.  
Assuming the setting in the above definition, $RB_{\Psi}(\psi|x) = \mathbb{E}_{\Pi(.|\psi)}(RB(\theta|x))$ 
and $\mathbb{E}_{\Pi_{\Psi}}(RB_{\Psi}(\psi|x)) = \mathbb{E}_{\Pi_{\Psi(.|x)}}(\frac{1}{RB_{\Psi}(\psi|x)}) = 1.$ 
where $\Pi(.|\psi)$ is the conditional prior given $\psi$, $\Pi$ is the prior and $\Pi(.|x)$ is the posterior. 
So, if $\psi$ is our parameter of interest, then the evidence for $\psi$ is the average evidence for parameter space w.r.t
the conditional prior given $\psi$, and the average evidence for parameter space w.r.t the prior is not informative or 
neutral. 

Another very natural and desired property for any valid measure of evidence is the following. 
Suppose $\varPsi = \{\psi_1, \psi_2\}$. Then $RB(\psi_1|x) < 1$ if and only if $RB(\psi_2|x) > 1$. 

\section{Strength of the Evidence}
So far the methods that we surveyed in the previous chapter were based on the idea that the evidence can be measured on
a universal scale. This does not seem to be correct and the subsequent flaws were examined in chapter 2. 
Hence in our context-dependant approach, the strength of the evidence will be determined in comparison to other possible 
values of the parameter in the parameter space. Perhaps the most informative and natural ingredient in this setting is the posterior 
distribution. Consider a simple yet instructive example.
\begin{example}
    
Suppose $\varPsi = \{\psi_1, \psi_2\}$.
When $RB(\psi_1|x) > 1$ and $\Pi_{\Psi}(\psi_1|x)$ is small, then there is evidence in favor of $\psi_1$ being the true 
parameter but our belief is weak. On the other hand, whenever $RB(\psi_1|x) > 1$ and $\Pi_{\Psi}(\psi_1|x)$ is small, i.e 
$\Pi_{\Psi}(\psi_2|x)$ is large, then, using the above proposition, we strongly believe that there is evidence against 
$\psi_1$ being the true value of the parameter. 
\end{example}
The above example outlines the motivation behind how we use the information in the posterior for calibration. Generalizing
the above for the case where $\#(\varPsi) > 2$ and continuous spaces, the following is used to measure the 
strength of the evidence. 
\begin{definition}{Strength of the evidence for $RB(\psi_0|x)$} \\
    When $RB(\psi_0|x) < 1$
    $$
    \Pi_{\Psi}(RB(\psi|x) \leq RB(\psi_0|x))
    $$    
    When $RB(\psi_0|x) > 1$
    $$
    \Pi_{\Psi}(RB(\psi|x) \geq RB(\psi_0|x))
    $$    
\end{definition}
There arguable many ways to measure the strength of the evidence. On can also use a small neighbourhood around $\psi_0$ 
in the posterior to determine the strength of the evidence in favor / against $\psi_0$. The above definition is a 
slightly modified version of the one presented in \parencite{evansMeasuringStatisticalEvidence2015}. I believe the above generalizes the above example more 
naturally and when $RB(\psi_0|x) > 1$, we are measuring our belief that the true parameter has evidence more than $\psi_0$ 
. While I do not have a formal justification to convince the reader that this might be a better measure for strength of 
the evidence, I do find this definition more plausible and future empirical experiments might shed more light. 

In the big picture, separating measurement of evidence from measuring its strength seems vital and natural. In the context
of parameter estimation, we need to supply our estimation with an accuracy, and this must hold for hypothesis
testing as well.
So such separation supplies us with both. Moreover, since relative belief induces a preference ordering on the parameter space,
estimation can be conducted naturally. The preferece ordering is as follows: 
$\psi_1$ is not strictly preferred to $\psi_2$, $\psi_1 \curlyeqprec \psi_2$, whenever $RB_{\Psi}(\psi_1|x) \leq 
RB_{\Psi}(\psi_2|x)$. 
Like any other preference ordering, objections might be raised. However, by keeping in mind that our parameters correspond
to some physical quantity, the relative belief preference ordering sounds plausible. Also, this preferece ordering is based on 
the arguably most important concept in statistics, namely evidence. Hence, by above total ordering, 
we use the Maximum Relative Belief Estimator (MRBE) as our estimate:
$$ 
\psi_{MRBE}(x) = \operatorname*{arg\,sup}_{\psi} RB_{\Psi}(\psi|x)
$$ 

Estimation can play a role in hypothesis assessment and resolve one of the main issues with p-values that 
has been mentioned in the previous chapter. Consider $(\mathbb{P}, \mathcal{A}, \Theta)$ and let $\Psi:\Theta \to \varPsi$. Let $H_0 = \Psi^{-1}\{\psi_0\}$ be the null hypothesis and suppose that there is evidence 
against the null because of large amount of data and an unmeaningful (in the context of a specific application) 
deviation from the null has been detected. A sensible method to settle this issue is to consider $|\psi_0 - \psi_{MRBE}|
<\delta$, where $\delta$ is supplied by the user and is a meaningful difference for a given application, to see 
whether the difference is meaningful. However, this issue automatically gets resolved as we first discretize the parameter space.

As argued before, an estimation should be supplied with a measure of accuracy. In parallel with other inference methods,
\begin{definition}{$\gamma$-relative belief region for $\psi$ is given by}
   $$
   \mathcal{C}_{\Psi,\gamma}(x) = \{\psi : Q_{\Psi}(RB_{\Psi}(\psi|x) \geq 1 - \gamma)\}
   $$ 
   Where $Q_{\Psi}$ is the posterior CDF. 
\end{definition}
Hence the "size" of the $\gamma$-relative belief region will determine the accuracy of our estimate and the notion of 
"size" needs to be meaningful for the given context. Moreover, for a desired $q > 0$, the following is 
\begin{definition}{q-Plausible region}
   $$
   pl_{\Psi,q}(x) = \{\psi:RB_{\Psi}(\psi|x) > q\}
   $$ 
   is the q-plausible region, and
   $$
   \Pi_{\Psi}(pl_{\Psi,q}(x)) 
   $$
   is the plausibility of $pl_{\Psi,q}(x)$. 
\end{definition}
Similar to the Bayes factor, the following expression for the relative belief ratio is very important. 
\begin{theorem}{General Savage-Dicky Ratio (Dickey 1971)}
$$
RB_{\Psi}(\psi|x) = \frac{m(x|\psi)}{m(x)} = \frac{m_{T}(T(x)|\psi)}{m_{T}(T(x))}
$$
where $m(.)$ is the prior predictive density, $m_{T}(.|\psi)$ is the conditional prior predictive given $\psi$, and 
$T$ is a minimal sufficient statistic for the model.
\end{theorem}

Relative belief regions and strength of the evidence has a number of properties which is common for Bayesian inference as 
well and can be found in \parencite{evansMeasuringStatisticalEvidence2015} p.122.

As seen in the Jefferys-Lindely paradox, the choice of prior is affecting the inference. In particular, diffuse priors 
might be introducing bias in favor of a particular hypothesis. Yet another benefit of defining statistical evidence is 
the ability to measure such bias a priori. The below characterization of bias is related to the idea of severe test \parencite{mayoSevereTestingBasic2006}.
\begin{definition}{Bias} \\ 
    Bias against $H_0 = \Psi^{-1}\{\psi_0\}$ is given by 
    $$
    M_{T}\Bigg(\frac{m_{T}(t|\psi_0)}{m_{T}(t)} \leq 1 \Big| \psi_0 \Bigg)
    $$
    And bias in favor of $H_0$ is 
    $$
    M_{T}\Bigg(\frac{m_{T}(t|\psi_0)}{m_{T}(t)} \leq 1 \Big| \psi' \Bigg)
    $$
    for any $\psi' \neq \psi.$  
\end{definition}
The interpretation is as follows. When $H_0$ is true, $M_{T}(\frac{m_{T}(t|\psi_0)}{m_{T}(t)} \leq 1 | \psi_0)$ being 
large indicates that our a priori belief for not finding evidence in favor of the null is large. Or in other words, there is 
a priori small probability of finding evidence in favor of $H_0$. Whenever $M_{T}(\frac{m_{T}(t|\psi_0)}{m_{T}(t)} \leq 1
| \psi')$ is small, and for values $\psi'$ with a meaningful difference from $\psi_0$, it indicates that the prior is 
biasing the evidence in favor of $\psi_0$. 

Before revisiting the Jefferys-Lindely paradox to illustrate the necessity of measuring bias, an important and interesting lemma is 
required where it demonstrates the relationship between the Bayes factor and the relative belief ratio.

\begin{lemma}
  For $A \subset \Theta = \varPsi$, $p > 0$ and $H_0 = \{\theta_0\}$, define the prior as $\Pi'(A) = p \delta_{H_0}(A) + 
  (1-p)\Pi(A)$ and let $\Psi:\Theta \to \varPsi$ be the identity map, where $\delta_{H_0}$ is the Dirac measure on 
  $\theta_0$ and $\Pi(\theta_0) = 0$. Then, 
  $$
  BF(H_0|x) = \frac{\pi(\theta_0 | x)}{\pi(\theta_0)} = RB(\theta_0 | x)
  $$
\end{lemma}
Hence the relative belief ratio and the Bayes factor agree in scenarios where $\theta_0$ is the value of the model parameter. However, it is 
important to note that relative belief does not adhere to an arbitrary choice of prior, but rather a logical development. 

\begin{example}{Jefferys-Lindely paradox revisited} \\ 
 Let $\Theta = \mathbb{R}$ and $\mathcal{M} = \{f_\theta : \theta \in \Theta\}$, where $f_\theta$ is the densify function
 for $N(\theta, 1)$. Moreover, let $x = (x_1,\dots x_n) \in \mathcal{X}$ be observed i.i.d and $T(X) = \bar{X}$ 
 be a minimum sufficient statistic. Hence, $T(X) \sim N(\theta, \frac{1}{n^{2}})$ 
 Our goal is to assess the hypothesis $H_0 = \{0\}$.  
 In relative belief setting, by the above lemma, 
 $$ 
 RB(0|x) = e^{- \frac{(n\sigma\bar{x})^2}{2(1+n\sigma^2)}}\sqrt{1 + n\sigma^{2}}
 $$
 This is same as the Bayes factor, but consider the behavior of the strength of the evidence for a very diffuse prior 
 as measured in \parencite{evansMeasuringStatisticalEvidence2015},
 $$
 \lim_{\sigma \to \infty}\Pi(RB(\theta|x) \leq RB(0|x)|x) = 2(1 - \Phi(|\bar{x}\sqrt{n}|))
 $$
 So p-value is really measuring the strength of the evidence in this scenario. 
 
 Consider a specific numerical example where
 $n = 50$, $\sigma^{2} = 400$ and $\bar{x}\sqrt{n} = 1.96$. Then $RB(H_0|x) = BF(H_0|x) = 20.72$ and by Jefferys' scale, this 
 is considered as strong evidence in favor of the null. However, the strength of the evidence is $0.05$ and so this 
 evidence is very weak. 
 Hence, it is concluded that large values of relative belief ratios, or Bayes factors by the lemma, does not indicate 
 strong evidence in favor of the null. Moreover, as prior becomes more and more diffuse, i.e. $\sigma \to \infty$, the evidence 
 in favor of the null becomes arbitrarily large. 
 Note that the problem of choosing the right $\sigma^{2}$ remains to be solved, and this is discussed in \parencite{evansMeasuringStatisticalEvidence2015} chapter 5. 

 Now consider the bias in favor calculation as follows,
 $$
 M_{T}(RB(0|x) \leq 1|\theta) = 1 - \Phi(c_n - \theta\sqrt{n}) + \Phi(-c_n - \theta\sqrt{n}),
 $$
 where $c_n = \sqrt{max(0, (1 + \frac{1}{(1 + n\sigma^{2})}log(1 + n\sigma^{2})))}.$ 
 Therefore, as $\sigma \to \infty$, the bias converges to $0$ for any $\theta$. So when $\theta = 0$, this is desirable and 
 otherwise, we are introducing bias in favor of the null. 

 Our estimation using relative belief preference ordering is $\psi_{MRBE}(x) = 20.72$; Assume that $20.72$ is meaningfully
 different from $H_0 = \{0\}$ for an application. Then $M_{T}(\frac{m_{T}(t|\psi_0)}{m_{T}(t)} \leq 1 | \psi_{MRBE}(x)) =
 0.12$. So we can suspect that, at least for $\psi'=\psi_{MBRE}(x)$, obtaining weak evidence is due to the bias in the 
 prior. Hence there is no a priori bias against $H_0$, but there is some in favor of it. 
 \\ \qed 
\end{example}
Such definition for bias has a very sensible and desired property, 
\begin{theorem}{Convergence of bias measure}
$$
\lim_{n \to \infty}M_{T_n}(\frac{m_{T_n}(t|\psi_0)}{m_{T_n}(t)} \leq 1 | \psi_0) = 0$$
$$ 
\lim_{n \to \infty}M_{T_n}(\frac{m_{T_n}(t|\psi_0)}{m_{T_n}(t)} \leq 1 | \psi') = 1
$$
Where $\psi' \neq \psi.$ 
\end{theorem}
Hence, in scenarios where sample size can be controlled, one can control bias and this can resolve the above paradox. 
However, this does not mean that bias calculation should be part of prior selection phase. In fact, priors should be 
elicited and then, for selected prior, calculations of bias should be performed to detect any possible issues. 
There are numerous optimality properties that relative belief posses and the reader is referred to \parencite{evansMeasuringStatisticalEvidence2015}.
An important distinction is that all those properties are proved for the finite case, since we believe that the correct 
behavior of an inference method in the finite case is sufficient, as argued in the first chapter.

\section*{Conclusion}
This short text tried to establish a big picture of what statistics is about and how an ideal inference method should 
behave. Moreover, by examining shortcomings of some of the currently used methods for measuring evidence and utilizing some intuitive principles, we motivated 
\textit{relative belief ratio} as the primary method of characterizing statistical evidence. Number of topics has been omitted for the interest of this text and the reader 
is strongly advised to refer to \parencite{evansMeasuringStatisticalEvidence2015} as the primary source for further readings of the subject. Further in chapter 4, optimality results are stated and in chapter 5, discussions regarding choosing and checking the ingredients of 
the statistical problem is conducted. 
Moreover, for a  
detailed comparison of the relative belief ratio and the Bayes factor, \parencite{al-labadiHowMeasureEvidence2024} should be consulted. 
    
\printbibliography

\end{document}